\documentclass[flushrt]{emulateapj}

\usepackage{subfigure} 
\usepackage{epsfig}

\newcommand{\npar}{\par \vspace{2.3ex plus 0.3ex minus 0.3ex}} 

\begin{document}

\title{A bright, spatially extended lensed galaxy at z = 1.7 \\ behind the cluster RCS2 032727-132623. \footnotemark[*]}

\author{Eva Wuyts\altaffilmark{1,2}, L. Felipe Barrientos\altaffilmark{3}, Michael D. Gladders\altaffilmark{1,2,4}, Keren Sharon\altaffilmark{2}, Matthew B. Bayliss\altaffilmark{1,2}, Mauricio Carrasco\altaffilmark{3}, David Gilbank\altaffilmark{5}, H.~K.~C. Yee\altaffilmark{6}, Benjamin P. Koester\altaffilmark{1,2},Roberto Mu{\~n}oz\altaffilmark{7}}

\footnotetext[*]{Based in part on observations collected at the European Southern Observatory, Chile; the 6.5\,m Magellan telescopes located at Las Campanas Observatory, Chile; the Southern Astrophysical Research 4.1\,m telescope, a collaboration between CNPq-Brazil, NOAO, UNC and MSU; the 3.5\,m Apache Point Observatory telescope in New Mexico, which is owned and operated by the Astrophysical Research Consortium; and the Canada-France-Hawaii Telescope (CFHT) on Mauna Kea, which is operated by the National Research Council of Canada, the Institut National des Sciences de l'Univers of the Centre National de la Recherche Scientifique of France, and the University of Hawaii.}
\altaffiltext{1}{Department of Astronomy and Astrophysics, University of Chicago, 5640 S. Ellis Av., Chicago, IL 60637}
\altaffiltext{2}{Kavli Institute for Cosmological Physics, University of Chicago, 5640 South Ellis Avenue, Chicago, IL 60637}
\altaffiltext{3}{Departamento de Astronom\'ia y Astrof\'isica, Pontificia Universidad Cat\'olica de Chile, Avda. Vicu\~na Mackenna 4860, Casilla 306, Santiago 22, Chile}
\altaffiltext{4}{Visiting Associate, Observatories of the Carnegie Institution of Washington, Pasadena, CA 91101}
\altaffiltext{5}{Department of Physics and Astronomy, University of Waterloo, Waterloo, Ontario, N2L 3G1, Canada}
\altaffiltext{6}{Department of Astronomy and Astrophysics, University of Toronto, 50 St. George St., Toronto, Ontario, M5S 3H4, Canada}
\altaffiltext{7}{Departamento de F\'isica y Astronom\'ia, Universidad de Valapara\'iso, Avda. Gran Breta\~na 1111, Valpara\'iso, Chile}

\begin{abstract}
We present the discovery of an extremely bright and extended lensed source from the second Red Sequence Cluster Survey (RCS2). RCSGA 032727-132609 is spectroscopically confirmed as a giant arc and counter-image of a background galaxy at $z=1.701$, strongly-lensed by the foreground galaxy cluster RCS2 032727-132623 at $z=0.564$. The giant arc extends over $\sim 38$\,\arcsec and has an integrated $r$-band magnitude of 19.1, making it $\sim 20$ times larger and $\sim 3.5$ times brighter than the prototypical lensed galaxy MS1512-cB58. This is the brightest distant lensed galaxy in the Universe known to date. 
We have collected photometry in 9 bands, ranging from $u$ to $K_s$, which densely sample the rest-frame UV and optical light, including the age-sensitive 4000\AA\ break. A lens model is constructed for the system and results in a robust total magnification of $2.04 \pm 0.16$ for the counter-image; we estimate an average magnification of $17.2 \pm 1.4$ for the giant arc based on the relative physical scales of the arc and counter-image on the sky. Fits of single-component spectral energy distribution (SED) models to the photometry result in a moderately young age, $t=80\pm40$\,Myr, small amounts of dust, $E(B-V) \le 0.11$, and an exponentially declining star formation history with \textit{e}-folding time $\tau = 10-50$\,Myr. After correcting for the lensing magnification, we find a stellar mass of M$_* \sim 10^{10}$\,M$_\odot$ and a current star formation rate SFR$\le77$\,M$_\odot$\,yr$^{-1}$. Allowing for episodic star formation, an underlying old burst could contain up to twice the mass inferred from single-component modeling. 
RCSGA 032727-132609 is typical of the known population of star-forming galaxies near this redshift in terms of its age and stellar mass. Its large magnification and spatial extent provide a unique opportunity to study the physical properties of an individual high-redshift star-forming galaxy in great detail, opening up a new window to the process of galaxy evolution between $z=1.7$ and our local Universe.

\subjectheadings{galaxies: evolution, galaxies: high-redshift, gravitational lensing}                                    
\end{abstract}

\section{Introduction}
Significant progress has been made in recent years towards the study of the formation of galaxies and their evolution into the population of objects we observe around us today. The redshift range $1.0 \lesssim z \lesssim 3.0$ represents a crucial period in this process since the peak of star formation in the Universe is known to occur near $z \sim 2$ \citep{Blain:99, Chapman:03, Reddy:09}. Through the development of pre-selection color criteria \citep{Adelberger:04, Steidel:04, Daddi:04a} and advances in the near-IR and UV spectroscopic capabilities of 8-10\,m class telescopes, growing samples of optically or near-IR selected galaxies are now spectroscopically confirmed in this redshift range and studied extensively (see e.g. Shapley et al. 2005; Erb et al. 2006; Kriek et al. 2008).
For a complete understanding of the process of galaxy formation, it is crucial to complement these statistical results based on larger samples with detailed study of the stellar populations and dynamics of individual objects. The majority of galaxies in the current samples are too faint for this purpose. The few galaxies that are bright enough to be studied individually represent outliers drawn from the extreme bright tail of the luminosity function, and are therefore not necessarily representative of the bulk of the population. The ability to study the properties of faint, high-redshift galaxies is one of the main science drivers for the construction of 30\,m class telescopes. Until such instruments become available we can make a head start using gravitational lensing to increase the power of the current generation of telescopes. 

The lensing magnification induced by foreground galaxy clusters and individual galaxies has been succesfully used to identify galaxies out to $z \sim 10$, opening up new windows into the very distant Universe \citep{Richard:08, Bouwens:09}. At more moderate redshifts it brings individual galaxies from the samples at $1.0 \lesssim z \lesssim 3.0$ to a flux level amenable for extensive follow-up observations at various wavelengths. The first and most notable example in this class is MS1512-cB58, a Lyman Break Galaxy (LBG) at $z = 2.73$ \citep{Yee:96, Ellingson:96}, found to have a lensing magnification of $\sim$ 30 \citep{Williams:96, Seitz:98}. It has been studied extensively since its discovery and provides a wealth of information on the stellar population and dynamics of a young star-forming galaxy at this redshift \citep{Pettini:00, Pettini:02, Siana:08}. Other examples of particularly bright lensed galaxies at similar redshifts include the `Cosmic Eye' at $z=3.07$ \citep{Smail:07,Coppin:07}, the `8 o'clock arc' at $z=2.73$ \citep{Allam:07, Finkelstein:09} and two strongly-lensed $z\sim3$ LBGs from the SDSS Giant Arcs Survey \citep{Koester:10}. These highly magnified sources currently represent the best places to study the individual properties of high-redshift star-forming galaxies in great detail. 

In this paper we present the discovery of a very bright and extended galaxy, RCSGA 032727-132609, spectroscopically confirmed at $z = 1.701$ and highly magnified by a foreground cluster at $z = 0.564$ from the second Red Sequence Cluster Survey (RCS2; D. Gilbank et al. 2010, in preparation). The RCS surveys were designed for the purpose of cluster finding via the identification of the linear color-magnitude relation present for early-type galaxies in clusters, known as the red-sequence technique \citep{Gladdersyee:00}. The RCS2 survey has imaged $\sim700$ square degrees of sky in $g$,$r$ and $z$ with the MegaCam camera at the Canada-France-Hawaii Telescope (CFHT) on Mauna Kea in 4, 8 and 6\,min exposures respectively. Data acquisition for the survey finished in 2008 and preliminary cluster catalogs over the entire area have been created and visually inspected for strong lensing signatures; the details of this search will be published elsewhere. The brightest and most obvious strong lensing system found in this search is RCSGA (RCS Giant Arc) 032727-132609. The system consists of a counter-image and a giant arc extending over $\sim 38$\,\arcsec at an Einstein-radius of $\sim 17.8$\,\arcsec, estimated from the distance between the brightest knot in the arc and the brightest cluster galaxy. The arc has an apparent magnitude of $r = 19.1$, making it $\sim 3.5$ times brighter than cB58 and the brightest distant lensed galaxy in the Universe known to date. Its large spatial extent provides unique opportunities to look inside a high-redshift galaxy and spatially resolve its substructure.
\npar
The paper is organized as follows. \S~\ref{sec:data} presents the multitude of photometric and spectroscopic data we have assembled on this system: broadband observations in 9 bands ranging from $u$ to $K_s$, a medium-resolution optical spectrum of the lensed galaxy to obtain the source redshift and redshift measurements of 49 cluster members to estimate the virial mass of the foreground cluster. \S~\ref{sec:phot} describes an innovative method used to obtain accurate photometry of the source. A lens model is constructed for the cluster in \S~\ref{sec:lensmodel}. We fit spectral energy distribution (SED) models to the broadband photometry of the source to explore its star formation history and stellar population parameters; the SED modeling procedure and results are discussed in \S~\ref{sec:sedall}. RCSGA 032727-132609 is compared to the known galaxy population at $z\sim 2$ in \S~\ref{sec:compall}. Throughout this paper we will assume $\Omega_M = 0.3$, $\Omega_\Lambda = 0.7$ and H$_0 = 70$\,km\,s$^{-1}$\,Mpc$^{-1}$. All magnitudes are quoted in the AB system.

\section{Observations and Data Reduction}
\label{sec:data}
\subsection{Imaging}
We have obtained imaging of RCSGA 032727-132609 in 9 bands ranging from $u$ to $K_s$ on 5 different telescopes. Details of these observations are given below and summarized in Table~\ref{tab:data}.

\begin{deluxetable*}{ccccc}
\tablecolumns{5}
\tablewidth{0pc}
\tablecaption{Summary of Imaging Data \label{tab:data}}
\tablehead{
\colhead{Filter}& \colhead{Total Int.}  & \colhead{Seeing} & \colhead{Date(s)} & \colhead{Telescope, Instrument}\\
\colhead{}& \colhead{(seconds)}  & \colhead{(arcsec)} & \colhead{Observed} & \colhead{}
}
\startdata
$u$&	      3600&	0.85&   2008-10-29&	SOAR, SOI  \\
$B$&	      738&	0.84&	2006-08-17&	VLT, FORS2\\
$g$&	      240&	0.74&	2005-11-04&	CFHT, MegaCam\\     
$r$&	      480&	0.64&	2005-11-04&	CFHT, MegaCam  \\   
$I$&	      480&	0.66&	2006-08-17&	VLT, FORS2\\
$z$&	      1080&	0.73&	2009-10-21&	Magellan Baade, IMACS f/2\\	
$J$&	      2160&	0.57&	2009-10-19&	Magellan Baade, PANIC\\
$H$&	      7200&     0.96&	2009-02-06&	Apache Point 3.5m, NICFPS\\
 &		&	&	2009-11-28&	\\
$K_s$&	      2160&	0.51&	2009-10-19&	Magellan Baade, PANIC\\
\enddata
\end{deluxetable*}

The $u$-band observation is part of a larger program to survey $\sim$100 strong lensing galaxy clusters in order to identify lensed $u$-band dropouts and constrain the fraction of strongly lensed galaxies that lie at z$\gtrsim3$ (Bayliss et al. 2010 in preparation). We observed RCSGA 032727-132609 in the SDSS $u$-band at the 4.1\,m Southern Astrophysical Research (SOAR) Telescope in Chile with the SOAR Optical Imager (SOI) on 2008, October 29 in six exposures of 600\,s each. The data were reduced and stacked in IRAF \footnotemark[1] using procedures from the MSCRED package. 
\footnotetext[1]{IRAF (Image Reduction and Analysis Facility) is distributed by the National Optical Astronomy Observatories, which are operated by AURA, Inc., under cooperative agreement with the National Science Foundation.}

Pre-imaging with the FORS2 instrument at the Very Large Telescope (VLT) in Chile for the purpose of mask design for the spectroscopy described below, consisted of dithered integrations in the $I$- and $B$-bands totalling 480\,s and 738\,s respectively. The imaging in both filters was acquired on 2006, August 17.

The $g$- and $r$-band data come from the RCS2 dataset, as detailed above. The data were acquired in queue mode in semester 2005B using the prime focus MegaCam imager on CFHT. It is important to keep in mind that the $g$-filter used for MegaCam on CFHT is similar to, but subtly different (both in central wavelength and width) from its namesake in the Sloan Digital Sky Survey (see Gilbank et al. in preparation for further details). The processed images are used as delivered by the CFHT queue pipeline, although astrometric and photometric re-calibration is performed as part of the RCS2 catalog processing. Unlike other data used in our analysis these images are single exposures with no dithering; cosmic rays and chip defects - neither of which are significant - have been identified by visual inspection and removed by interpolation.

During a run in October 2009, the IMACS instrument and the PANIC near-IR camera on the Magellan I Baade Telescope were used for broadband imaging of RCSGA 032727-132609 in the $z$-, $J$- and $K_s$-bands. A total of 1080\,s of dithered $z$-band images were acquired in nine individual integrations with the IMACS f/2 spectrograph in imaging mode. The data were processed and stacked using standard IRAF tasks. Dithered observations in both $J$ and $K_s$ were acquired with the PANIC near-IR imager, totaling 2160\,s in both bands. The PANIC data were reduced using a custom pipeline built in IRAF with standard techniques of dark subtration, flat-fielding, and sky subtraction with iterated object masking. 

\begin{figure*}
\centering
\includegraphics[width=10cm]{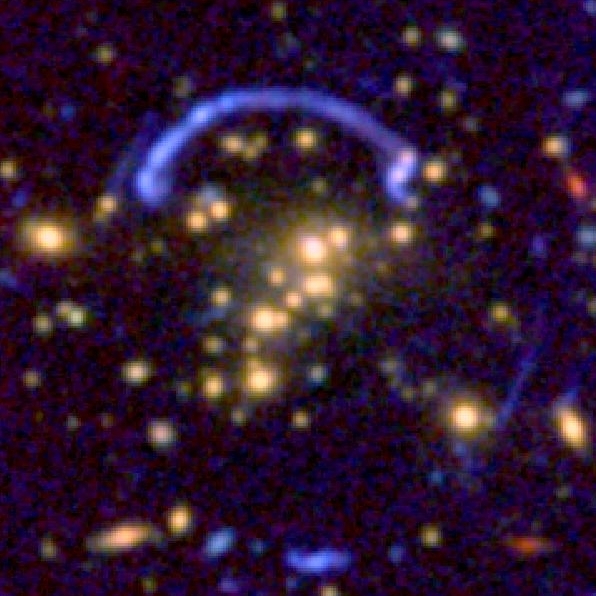}
\vspace{0.30in}
\caption{Composite color-image of $uBg$, $rIz$ and $JHK_s$ combined imaging data, ($1\times1$\,arcmin). All images were transformed to the pixel scale of the VLT $I$-band image and the PSFs were degraded to match those of the $H$-band, which has the worst seeing. North is up and East is left. \label{fig:color}}
\end{figure*}

\begin{figure*}
\centering
\includegraphics[width=10cm]{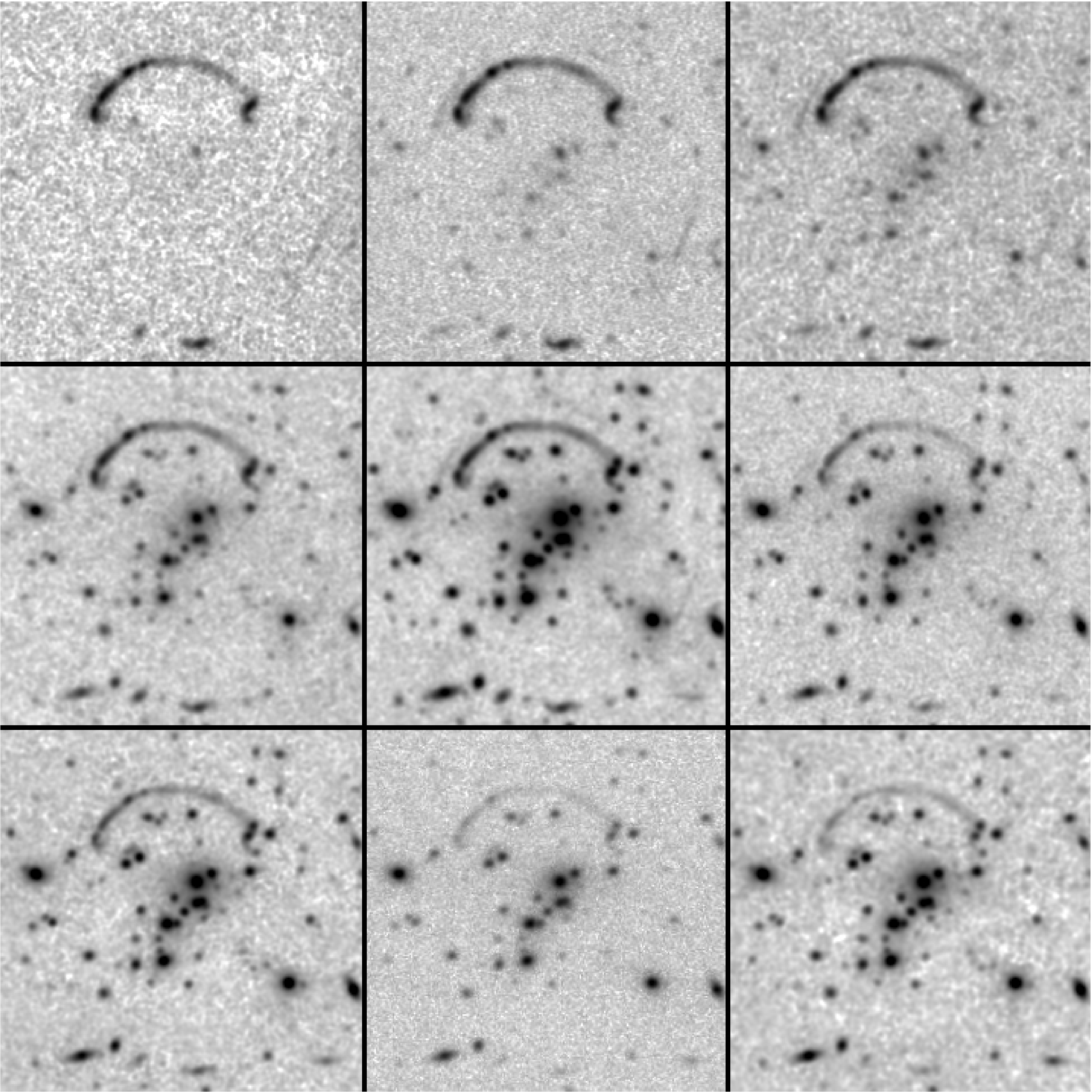}
\vspace{0.1in}
\caption{Gallery of $1\times1$\,arcmin images of RCSGA 032727-132609 in $u,g,B,r,I,z,J,H$ and $K_s$ (top left to bottom right). All images were transformed to the pixel scale of the VLT $I$-band image and the PSFs were degraded to match those of the $H$-band, which has the worst seeing. North is up and East is left. \label{fig:bw}}.
\end{figure*}

Finally, an $H$-band image was constructed from 7200\,s of dithered observations taken on 2009, February 6 and November 28, with the 3.5\,m telescope at Apache Point Observatory (APO) in New Mexico. Additional short exposure $J$- and $K_s$-band imaging was collected for calibration purposes using the same method. We used a pipeline of standard IRAF tasks to dark-subtract and stack the dithered images.
\npar
Figure~\ref{fig:color} shows a composite color-image of RCSGA 032727-132609 constructed from all 9 bands, combining $uBg$, $rIz$ and $JHK_s$ individual composites. Postage stamp cutouts of all 9 photometry bands are shown in Figure~\ref{fig:bw}. The west end of the arc is complicated by two cluster members that fall on top of it (see \S~\ref{sec:phot} and Figure~\ref{fig:galfit}); the $u$-band data obviously shows this region to be part of the giant arc.

\subsection{VLT Spectroscopy}
\label{sec:spec}
\begin{figure}[h]
\centering
\includegraphics[width=9cm]{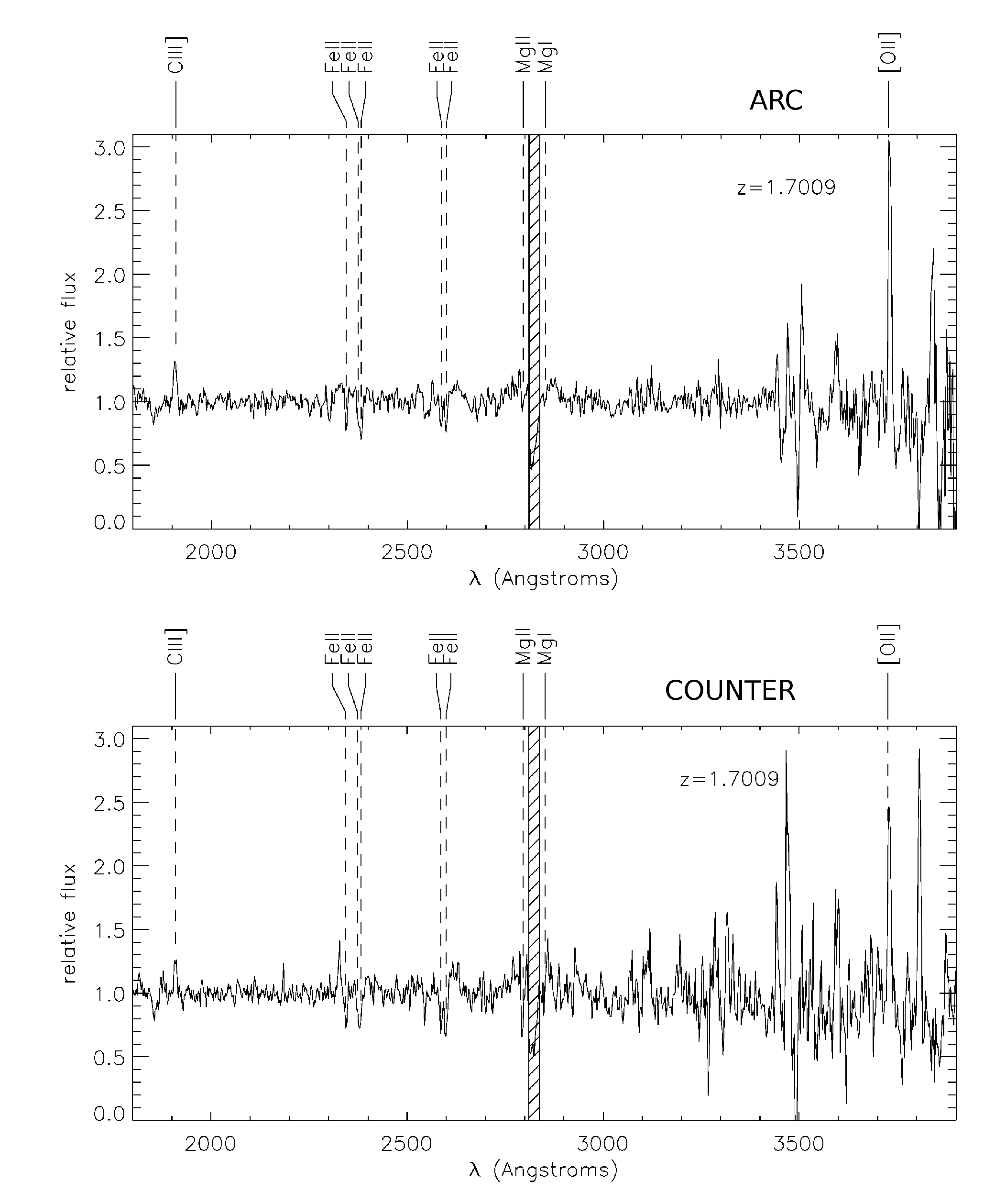}
\vspace{0.05in}
\caption{Normalized stacked VLT spectra of the arc (top) and the counter-image (bottom) at rest-frame wavelengths. Important spectral lines are identified.  \label{fig:spectrum}}
\end{figure}

The Very Large Telescope (VLT) in Chile was used to measure the redshift of the lensed background galaxy. A total integration time of 2 hours was obtained with FORS2 with a multi-slit mask in service mode on the nights  of December 27, 2006 and January 13 and 27, 2007. The setup used the GRIS150I disperser and GG435 filter, providing a wavelength coverage from 4450 to 10500\AA\ with a resolution of 3.45\AA\,pixel$^{-1}$. The system was binned to 2 pixels both spatially and spectrally. Spectra were taken of several locations in the main arc, a likely counter-image, and a few possible cluster galaxies selected on the basis of $B-I$ color as measured from the mask design pre-imaging. Reductions were carried out with the ESO package esorex\footnotemark[1] and wavelength calibration was done by comparison to standard He+Ne+Ar lamp observations. Individual images were sky-line subtracted and then combined into a single image to eliminate cosmic rays. Sky-subtraction is complicated by the complexity of the object placement on the slits, as this precluded along-slit dithering. There are thus some significant fringing residuals at redder wavelengths. Nevertheless, the lensed source redshift is obvious due to strong absorption and emission lines; these spectra also confirm the identity of the counter-image. The stacked 1D spectra for the arc and  counter-image are shown in Figure~\ref{fig:spectrum}. The source redshift is 1.7009 $\pm$ 0.0008, established from the FeII and MgII lines in absorption and the [OII]$\lambda3727$ and the CIII]$\lambda1909$ lines in emission.
\footnotetext[1]{http://www.eso.org/sci/data-processing/software/pipelines/fors}
\npar
One prominent feature in the spectrum is the CIII]$\lambda1909$ nebular emission line, a collisionally excited, semi-forbidden transition often present in local starbursts. It is found to be stronger in starbursts of lower metallicity \citep{Heckman:98}, most likely due to a decrease in the nebular electron temperature of higher metallicity gas, which causes more of the nebular cooling from collisionally excited lines to occur in the infrared rather than the UV. At higher redshifts this line is usually too weak to be identified in individual spectra, but it has been detected in composite spectra of star-forming galaxies at $z \sim 2$ \citep{Erb:06, Halliday:08} and in a composite LBG spectrum at $z\sim3$ \citep{Shapley:03}. The inverse relation between the strength of CIII] nebular emission and the metallicity of the galaxy remains valid at these higher redshifts. It is interesting to note that the CIII] line was not detected in the UV spectrum of cB58 \citep{Pettini:00,Pettini:02}. This emission line can also be used in combination with either Ly$\alpha$ or CIV line intensities as an indicator of AGN contribution \citep{Shapley:03}.

\subsection{Cluster Velocity Dispersion}
\label{sec:veldisp}
\begin{figure}
\centering
\includegraphics[width=9cm]{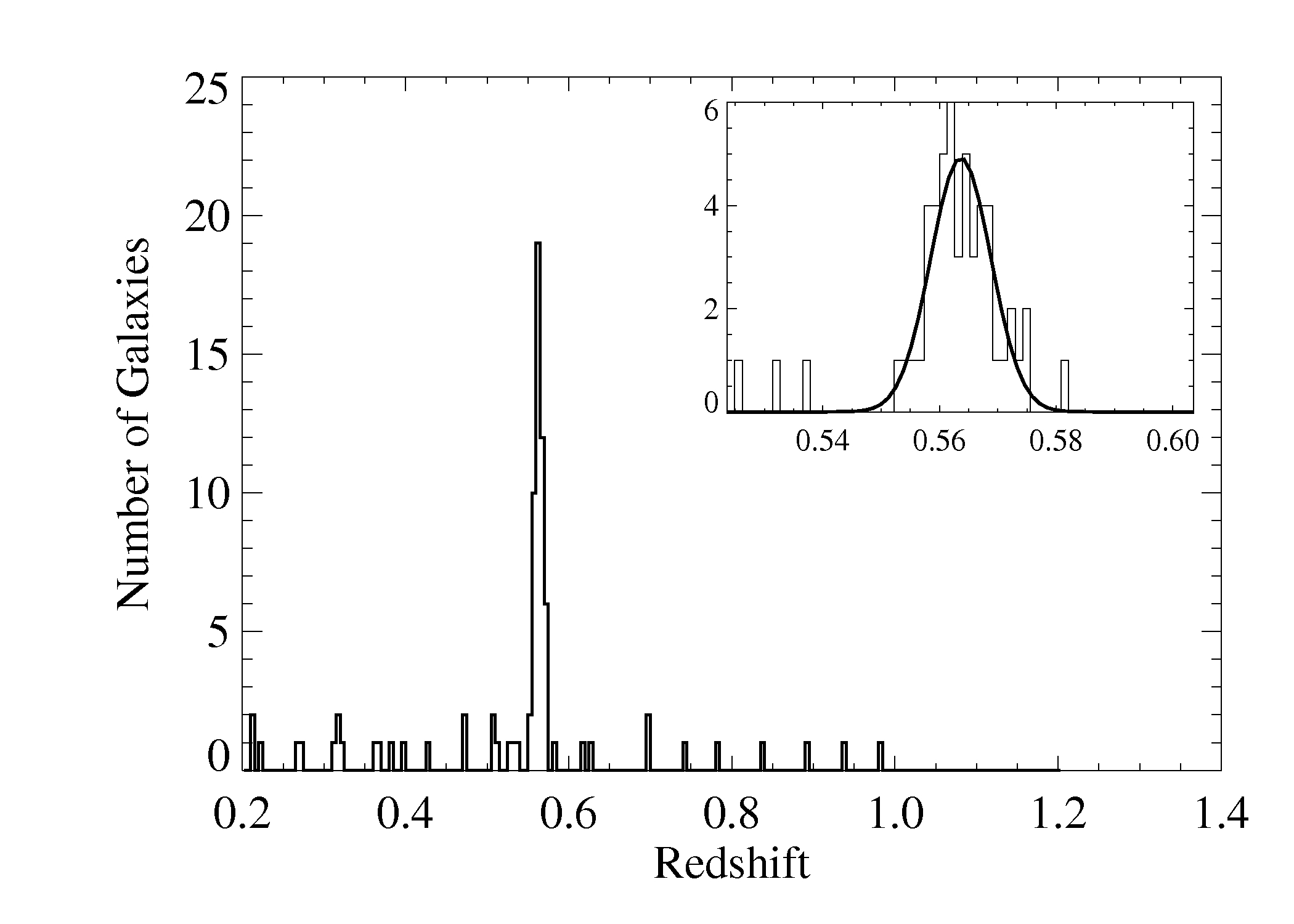}
\caption{The main panel shows a histogram of redshifts binned at a redshift interval of 0.005, illustrating the dominance of the cluster in redshift space. The inset shows the velocity distribution relative to the cluster mean redshift, binned at a velocity interval of 250 km~sec$^{-1}$ (rest-frame), with the best fitting gaussian distribution overplotted. \label{fig:veldisp}}
\end{figure}

The cluster RCS2 032727-132623 was observed spectroscopically with the GISMO instrument (Gladders et al., in preparation) on the Magellan I Baade Telescope on January 25, 2009, using the 150 l/mm grating and the WB5650-9200 spectroscopic filter. A single mask covering a roughly circular area on the sky with a radius $\sim 1.75$\,arcmin, which corresponds to a physical radius of $\sim 0.9$\,Mpc at the redshift of the cluster, was observed in three integrations of 2400\,s each (7200\,s total). Data were reduced using the COSMOS pipeline\footnotemark[2]. These data yielded a total of 66 reliable redshifts, of which 46 are cluster members. Three additional cluster members have measured redshifts from the VLT spectroscopy discussed above. 
Figure~\ref{fig:veldisp} shows a histogram of all measured redshifts in this field. No other significant structures are detected in redshift space to the limits of these data; given the tight centering of these data on the cluster core this is not surprising, though it does confirm that no other cluster-scale halos at $z<1$ contribute to the strong lensing along the line of sight. The cluster has a measured redshift of 0.5637$\pm$0.0007. The best fitting velocity dispersion is $988\pm122$\,km\,s$^{-1}$ in the rest-frame, corresponding to a $M_{200} \sim 1.1 \times 10^{15}$\,M$_\odot$ cluster \citep{yee03}, in accord with the large arc radius observed. There is no evidence for substructure in the cluster in velocity space. 
\footnotetext[2]{http://obs.carnegiescience.edu/Code/cosmos}

\vspace{60pt}
\section{Photometry}
\label{sec:phot}
\begin{figure}[h]
\plotone{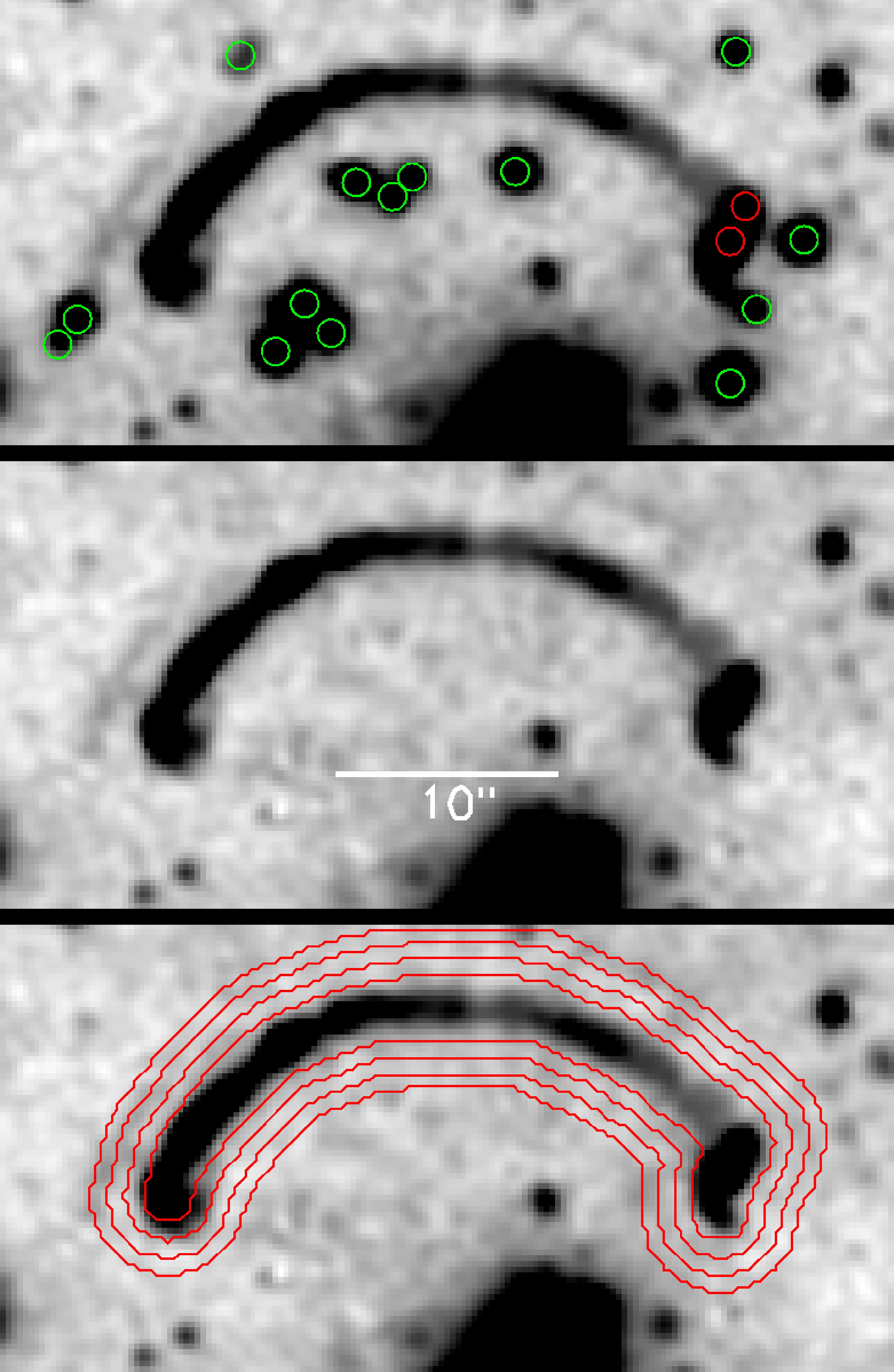}
\vspace{0.25in}
\caption{\textit{Top:} Cutout of PSF-matched $I$-band; circles show galaxies that need to be subtracted - the red circles correspond to the 2 cluster members on top of the arc. \textit{Middle:} Final subtracted $I$-band image. \textit{Bottom:} Final $I$-band image with arc apertures. \label{fig:galfit}}
\end{figure}

We have developed a self-consistent, complex IDL-code to obtain accurate magnitudes for extended sources in a series of broadband images with differing point spread functions (PSF). All images are first transformed to the orientation and scale of the VLT $I$-band image. An empirical, normalized PSF is created for each image based on a well-defined, isolated and non-saturated reference star. We match the PSF of each frame to the $H$-band image (which presents the worst seeing over our sample of 9 bands) by convolving it with an appropriate gaussian. This sequence of PSF-matching ensures that we measure the light in the same physical region of the source in all bands. This is a crucial requirement for meaningful modeling of the spectral energy distribution at a later stage.  The PSFs of these data are sufficiently similar that convolution with a simple gaussian provides adequate matching between images. Object apertures are created by tracing a curve along the extended source and convolving it with the $H$-band PSF. A series of apertures of increasing radial extent are defined as isophotes of this convolution. The apertures are very much non-circular, especially for the arc, and described by equivalent radii based on circular apertures that extend to the same isophotes. After a detailed sequence of sky subtraction and outlier masking, final magnitudes are measured at an equivalent radius of twice the FWHM of the $H$-band image, 7.6 pixels or $1\farcs92$, and aperture-corrected to an equivalent radius of $6''$ based on the curve of growth of the PSF reference star. 
\npar
Lensed background sources are often found close to the center of the foreground galaxy cluster. In these generally crowded environments, problems arise when cluster members lie too close to the background galaxy and fall inside the measurement apertures. We use the GALFIT package \citep{Peng:02} to fit a Sersic profile to these galaxies in the reference $I$-band image. This model is convolved and scaled as needed, and subtracted from each PSF-matched image. Figure~\ref{fig:galfit} shows the $I$-band image before and after this model subtraction, and overlayed with the arc apertures used for the final measurement.
In the case of RCSGA 032727-132609, we are presented with the unfortunate and complex situation where 2 cluster members fall on top of the giant arc. They are circled in red in the top panel of Figure~\ref{fig:galfit}. Evidently their flux contribution has to be subtracted from the arc magnitude, but in this case the appropriate scaling can not be estimated easily. We create a color image for each frame by subtracting a scaled version of the $r$-band to make the arc vanish. The GALFIT models of the overlapping cluster members are then used to subtract any remaining positive flux at their positions and the original frame is restored by adding the scaled $r$-band back in. This process is unnecessary for the $u$-, $B$- and $g$-bands, were the contribution of the cluster members is negligible, and becomes more important for the longer wavelengths. We confirm the subtraction by comparing the brightness of the affected part of the arc to the brightness of other parts (an advantage of large spatial coverage). Through magnitude measurements of visually obvious over- and undersubtracted frames, we estimate the uncertainty which arises from this subtraction: 0.01\,mag for $r$, $I$ and $z$; 0.02\,mag for $J$; 0.03\,mag for $H$; 0.04\,mag for $K_s$. This is incorporated into the final photometric uncertainties. 

\npar
The calibration of the images is handled carefully to ensure an accurate magnitude measurement of our object of interest. We calibrate the $g$-, $r$- and $z$-bands against the RCS2 catalog, which itself has been calibrated to match the SDSS stellar locus (Gilbank et al. in preparation). There is a 0.05\,mag absolute calibration uncertainty associated with the RCS2 photometry. The $B$- and $I$-bands are calibrated from the $grz$-photometry using transformations by \cite{Jordi:06} and \cite{Chonisgaskell:08}. The $u$-band calibration follows from the $gr$-photometry via a general relation derived from SDSS data. The near-IR $J$-, $H$- and $K_s$-bands are calibrated against the 2 Micron All Sky Survey (2MASS). The small field of view of the Magellan $J$ and $K_s$ images does not include enough 2MASS stars for a robust calibration, so we use the larger APO images to pick up enough 2MASS objects and link the APO zeropoint to the Magellan zeropoint based on a number of fainter point sources. Each star used for zeropoint calibration undergoes the same procedure explained above for its flux measurement. This eliminates any differences between our measurement method and the surveys we are calibrating against and results in the most accurate zeropoint determination. We add the zeropoint uncertainty and the Poisson error from the flux measurement in quadrature to construct final photometric uncertainties. 
\npar
Final magnitudes are corrected for galactic extinction \citep{Schlegel:98} and presented in Table~\ref{tab:mag} for the main arc and counter-image seperately. The photometry of the smaller and more regularly-shaped counter-image does not require the careful masking of cluster galaxies inside the source aperture and its measurement resembles much more a typical photometry measurement. After taking into account the lensing magnifications, the counter-image can therefore be used as a consistency check for the photometry of the giant arc. The arc magnitudes are on average a factor of 2.34 brighter, which is consistent with the difference in their sizes due to the lensing
magnification as explained in \S~\ref{sec:lensmodel}.

\begin{deluxetable*}{lccccccccc}
\tabletypesize{\scriptsize}
\tablewidth{0pt}
\tablecaption{Final magnitudes for the arc (\textbf{A}) and counter-image (\textbf{C}).}
\tablehead{
\colhead{} & \colhead{$u$} & \colhead{$B$} & \colhead{$g$} & \colhead{$r$} &
\colhead{$I$} & \colhead{$z$} & \colhead{$J$} & \colhead{$H$} & \colhead{$K_s$}}
\startdata
\textbf{A}& 19.17$\pm$0.100& 19.14$\pm$0.070& 19.15$\pm$0.055& 19.07$\pm$0.065& 19.02$\pm$0.080& 18.90$\pm$0.065& 18.33$\pm$0.090& 18.36$\pm$0.075& 18.43$\pm$0.080 \\
\textbf{C}& 21.39$\pm$0.100& 21.26$\pm$0.075& 21.36$\pm$0.060& 21.29$\pm$0.060& 21.39$\pm$0.080& 21.22$\pm$0.070& 20.72$\pm$0.080& 20.83$\pm$0.140& 20.92$\pm$0.120 \\
\enddata
\label{tab:mag}
\end{deluxetable*}

\section{Lens model}
\label{sec:lensmodel}
The multiply-lensed system  RCSGA 032727-132609 is composed of a highly-stretched giant arc north of the center of the cluster, and a much less magnified counter-image south of it. The separation between the giant arc and the counter-image is $\sim$46\,\arcsec. The structure of the giant arc is slightly complicated by the lensing potential of two cluster-member galaxies at its west side (see Figure~\ref{fig:galfit}). The giant arc is highly magnified in the tangential direction, revealing substructure that is not evident in the counter-image. With the resolution of the current imaging data we cannot uniquely identify the emission knots that form this substructure as multiply-imaged counterparts of the same emission region in the source. In constructing a lens model, we are therefore forced to make assumptions on the identity of these features. High resolution data may resolve some of the discrepancy and allow for a more detailed model in the future. 

\begin{figure}[h]
\centering
\includegraphics[width=9cm]{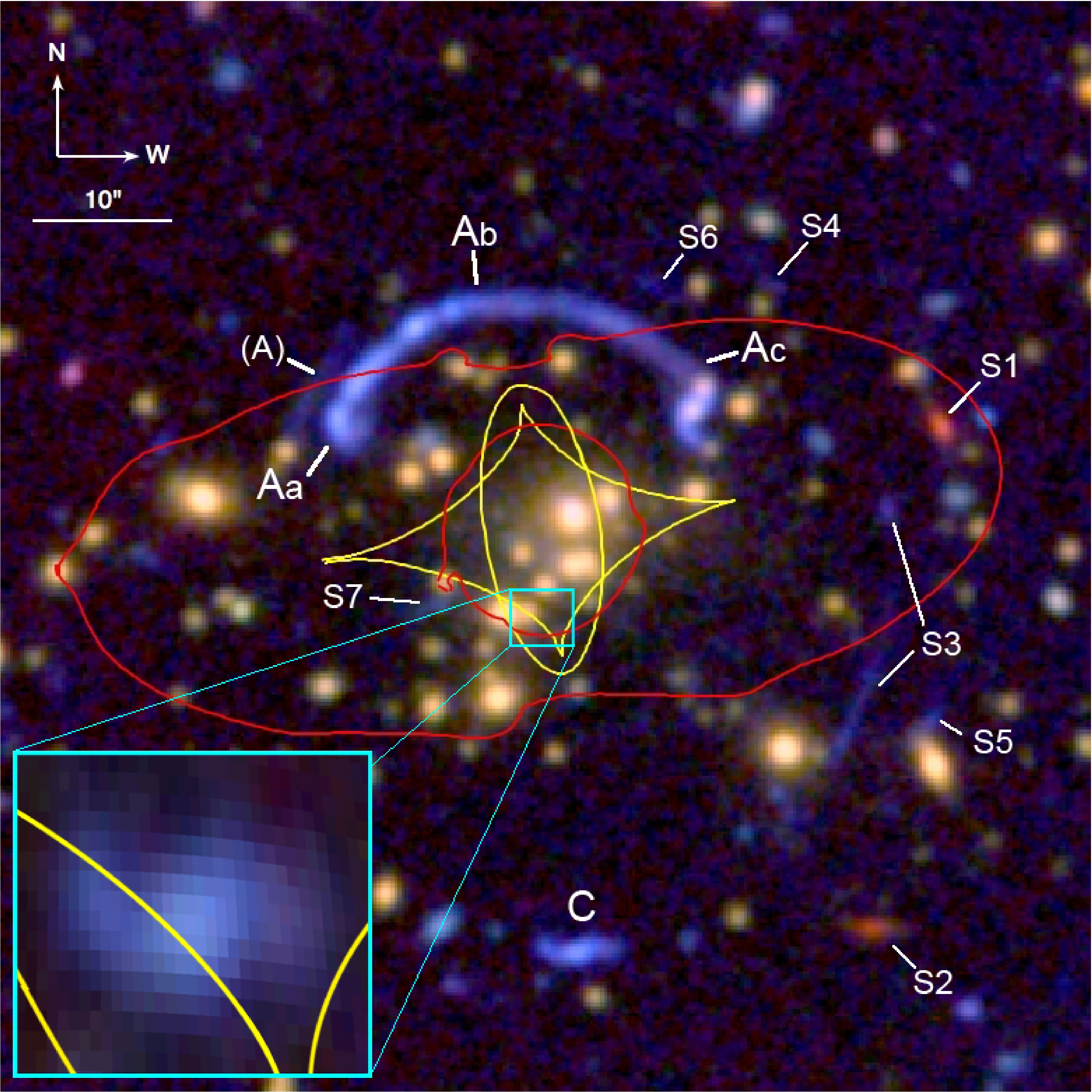}
\vspace{0.05in}
\caption{Strong lensing interpretation shown on a composite color image of $uBg$, $rIz$, and $JHK_s$. The critical curves of the best-fit lens model are overplotted in red: the outer curve is the tangential critical curve, and the inner one is the radial critical curve. The cuspy and oval yellow curves are the tangential and radial source-plane caustics, respectively. The three images that compose the giant arc RCSGA 032727-132609 are labeled with Aa, Ab and Ac. We identify the faint blue feature labeled with (A) as a part of the same source, but it is not included in the analysis presented in this paper. The counter-image is labeled with C. The inset shows a reconstruction of the source from the counter-image. Its location in the source plane is indicated by a cyan rectangle. We identify other likely lensed sources, and mark these candidates with S1-7. We note that S1 and S2 are $r$-dropout galaxies, S3 and S4 are likely a three-image system of the same source, and S7 is likely a radial arc. \label{fig:lensmodel}}
\end{figure}

The lens model is computed using the publicly-available software {\tt LENSTOOL} \citep{Jullo:07}, in an iterative manner. We start by representing the cluster halo by a single mass in the form of a pseudo-isothermal ellipsoidal mass distribution (PIEMD; Limousin et al. 2005). We use as constraints the redshift of the source, the coordinates of the counter-image, and a series of coordinates tracing the giant arc. The best-fit model is found through Markov Chain Monte Carlo (MCMC) minimization in the source plane. The resulting first-order model is not a satisfactory fit to the data, but it indicates that the giant arc is composed of three images of the source. As a second step, we break the giant arc into three different parts, using the observed symmetry of the substructure and guided by the location of the critical curves of the initial model. In our final model, we add the lensing contributions of the 30 brightest cluster-member galaxies, each represented by a PIEMD  with parameters that follow the observed properties of the galaxies through scaling relations (see Limousin et al. 2007 for further description of the scaling relations). Cluster members are selected via their location relative to the cluster red sequence in a color-magnitude diagram. We let all the parameters of the cluster halo vary in the MCMC minimization, but introduce priors on the velocity dispersion following the observed value (see \S~\ref{sec:veldisp}). The parameters of the best-fit model are enumerated in Table~\ref{table:param}, and the corresponding critical curves are shown in Figure~\ref{fig:lensmodel}.

\begin{deluxetable*}{lccccccc} 
\tablecolumns{8} 
\tablecaption{Best-fit lens model parameters \label{table:param}} 
\tablehead{\colhead{Mass}   & 
            \colhead{RA}     & 
            \colhead{Dec}    & 
            \colhead{$e$}    & 
            \colhead{$\theta$}       & 
            \colhead{$r_{\rm core}$} &  
            \colhead{$r_{\rm cut}$}  &  
            \colhead{$\sigma_0$}\\ 
            \colhead{}   & 
            \colhead{(h~m~s $\pm ''$)}     & 
            \colhead{($^\circ$~$'$~$''$ $\pm ''$)}     & 
            \colhead{}    & 
            \colhead{(deg)}       & 
            \colhead{(kpc)} &  
            \colhead{(kpc)}  &  
            \colhead{(km s$^{-1}$)}  } 
\startdata 
Cluster (PIEMD) & 03:27:27.310 $\pm {0.2}$ &-13:26:23.65 $^{+0.5}_{-0.3}$ &0.62 $^{+0.07}_{-0.06}$ &8.6 $\pm{0.4}$ &5.3 $\pm{1.2}$ &281 $^{+13}_{-115}$ &1094 $^{+ 7}_{-21}$ \\ 
L* galaxy (PIEMD) & \nodata & \nodata & \nodata & \nodata &  [0.15]  &     5.2 $^{+0.7}_{-3.2}$ &  87 $^{+32}_{-29}$ \\ 
\enddata 
 \tablecomments{The ellipticity is expressed as $e=(a^2-b^2)/(a^2+b^2)$. $\theta$ is measured north of West. Error bars correspond to 1$\sigma$ confidence levels as inferred from the MCMC optimization. Values in square brackets are not optimized. The location and the ellipticity of the matter clumps associated with the cluster galaxies were kept fixed according to the light distribution.}
\end{deluxetable*}

\npar
In order to scale the measured brightness back to the source plane and estimate the luminosity of the source, we need to know the average magnification of the source due to lensing. We estimate the magnification uncertainty through a Monte-Carlo simulation, in which we compute many lens models, in each one drawing a set of model parameters from steps in the MCMC that are within [$\chi^2_{min}$,$\chi^2_{min}+2$] of the best-fit model. The average magnification of the counter-image and its uncertainty as predicted by these models is $2.04 \pm 0.16$. 
The magnification of the giant arc is highly sensitive to its location relative to the critical curve in the image plane, and consequently very sensitive to the details of the model. Areas of the arc that lie closer to the critical curves are more highly magnified. The resolution of the current imaging data is insufficient to obtain separate magnifications for the 3 merged images that make up the giant arc or for any of the visible substructure. We therefore estimate an average magnification for the whole arc from the more robust magnification estimate of the counter-image and the relative sizes of the arc and counter-image on the sky. We count the number of pixels above a brightness treshold in the $I$-band data for both; for a range of surface brightness tresholds $\mu$ between 24.32 and 25.07\,mag arcsec$^{-2}$, we find a relative size ratio of $8.45\pm0.15$. After multiplying this by the magnification of the counter-image, we estimate the average magnification of the giant arc to be $17.2 \pm 1.4$.

\section{Stellar energy distributions}
\label{sec:sedall}
\subsection{Modeling procedure}
\label{sec:sedmodel}
The spectral energy distribution (SED) of a galaxy is governed by its star formation and metal enrichment history in conjunction with its current stage of stellar evolution and dust attenuation and as such contains information on the galaxy's stellar mass, age, star formation history, metallicity and dust content. It is unfortunately not an easy task to extract these parameters with a high degree of certainty from an observed SED, a problem which gets increasingly worse towards higher redshifts. Recent studies have shown that the choice of stellar evolution prescription is the largest systematic uncertainty in the interpretation of a galaxy's SED, overshadowing the random errors which arise from the photometric uncertainties. It is not a goal of this paper to conduct a complete comparison (see e.g. Maraston 2005; Conroy \& Gunn 2009; Muzzin et al. 2009; Wuyts et al. 2009), but it is important to understand the main aspects. One of the major challenges remains the treatment of thermally pulsating asymptotic giant branch (TP-AGB) stars, which contribute significantly to the restframe near-IR emission for galaxies in the age range $0.2 \le t \le 2$\,Gyr \citep{Maraston:05}. Since the near-IR traces a galaxy's old stellar population, this is a key issue in the study of stellar masses. The updated version of the stellar evolution models by Charlot \& Bruzual (hereafter CB07, kindly made available by the authors) includes a revised treatment of TP-AGB stars and brings them into closer agreement with the other widely used class of models by \cite{Maraston:05}. We use the models based on a Chabrier initial mass function \citep{Chabrier:03}. It is well established that a Salpeter IMF \citep{Salpeter:55} overpredicts the number of stars less massive than 1\,M$_\odot$, increasing the inferred stellar masses and star formation rates by a factor of $\sim$ 1.5-2 (see e.g. Papovich et al. 2001). We investigate the influence of a Salpeter IMF for RCSGA 032727-132609 in \S~\ref{sec:comp} to allow comparison with other studies. The dust extinction is governed by the Calzetti extinction law \citep{Calzetti:00}, which has been derived for local starbursts. The validity of this extinction law at high redshift has yet to be tested adequately. It has been suggested that the Calzetti law overpredicts dust extinction for young LBGs (ages $\lesssim 100$\,Myr) when compared to other diagnostics \citep{Reddy:06b, Siana:08, Siana:09, Reddy:10}. \cite{Muzzin:09} find a 10-20\% systematic effect on stellar population parameters from the choice of dust extinction law, small compared to the systematic errors connected to the choice of stellar evolution model.

Once the model set has been chosen, we have to worry about the internal degeneracies of a spectral energy distribution. Age, metallicity and dust all tend to affect a galaxy's spectrum in similar ways; the SED will overall be redder for a more metal rich stellar population, an older stellar population or more dust extinction. To decrease the uncertainties of the important stellar population parameters such as stellar mass and age which originate from this degeneracy, the metallicity is often treated as a known parameter. We use solar metallicity as a default choice, which is also most suitable for comparison with the literature. Lower metallicities are investigated in \S~\ref{sec:met}. The remaining degeneracy between the age of a stellar population and its reddening can be broken if near-IR photometry is included in the SED to cover the age sensitive 4000\AA\ break.
\npar
We use an updated version of the code \textit{Hyperz} \citep{Bolzonella:00}, kindly made available to us by M. Bolzonella, which performs SED fitting at a fixed spectroscopic redshift. The code matches the observed broadband SED to a set of template spectra through a maximum-likelihood procedure, quantifying the goodness of fit via $\chi^2$-statistics. It is important to note that the code does not interpolate on the template grids; the input template set must be densely populated. We use the solar metallicity templates from CB07 for a range of exponentially declining star formation histories of the form {\it SFR}$(t) \sim \exp(-t/\tau)$, with \textit{e}-folding times $\tau$ = 0.01, 0.05, 0.1, 0.2, 0.5, 1 and 2\,Gyr, as well as continuous star formation models (CSF). For the main analysis, we do not consider more complex star formation histories such as multiple component models with several short-duration bursts, since the data do not place strong constraints on even the single-component models. We only briefly investigate an extreme two-component model in the next section to determine an upper limit on the stellar mass from a maximally old underlying burst. Ages are allowed to vary between 0.1\,Myr and the age of the Universe at $z=1.701$ and extinction is constrained to be $E(B-V) \le 0.75$. 

The procedure outputs the best-fit spectral energy distribution and the corresponding stellar population parameters. Independent estimates are obtained for the arc and the counter-image and can be used as an internal consistency check of the photometry as well as the SED modeling procedure. The age $t$, reddening factor $E(B-V)$ and star formation history $\tau$ depend solely on the galaxy colors, and should be similar for both images of the same background galaxy. The stellar mass and current star formation rate have an additional dependence on the lensing magnification and are corrected for the magnification factors obtained in \S~\ref{sec:lensmodel}. The uncertainties related to the lens model need to be taken into account when comparisons are made for these two parameters.

\subsection{Results}
\label{sec:sedresults}
\begin{figure}[h]
\centering
\includegraphics[width=9cm]{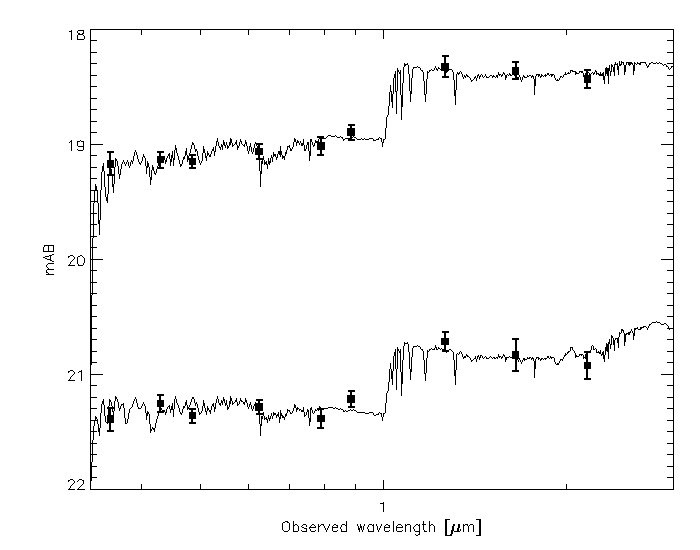}
\vspace{0.05in}
\caption{Best fit stellar energy distributions for the arc and counter-image. Photometry in $u,B,g,r,I,z,J,H$ and $K_s$-bands is overplotted as datapoints with $1\sigma$ errorbars. \label{fig:seds}}
\end{figure}

Figure~\ref{fig:seds} shows the best-fit SEDs for the arc and counter-image found from the set of models described above. The fits are good, with a reduced $\chi^2$ of 0.87 and 0.89 for the arc and counter-image respectively. It is important to supplement the best-fit stellar population parameters with parameter confidence intervals allowed by the photometric uncertainties. We create 1000 fake realizations of the observed SED by perturbing each broadband magnitude measurement in a manner consistent with its errorbars. This set of fake SEDs is fit in exactly the same manner as described above for the observed SED; bad fits with $\chi^2 > 3.0$ are excluded and the median and standard deviation are deduced for each stellar population parameter. The results are listed in Table~\ref{tab:sed}. The stellar mass and current star formation rate have been corrected for the magnification factors of 17.2$\pm$1.4 for the arc and 2.04$\pm$0.16 for the counter-image. The correspondence between the stellar population parameters independently inferred from the arc and counter-image is encouraging. 

The stellar mass is one of the most robust parameters that can be determined from SED modeling, least affected by intrinsic uncertainties related to the choice of stellar evolution model. However, there are other effects to be taken into account. First of all, the stellar mass reported from SED modeling corresponds to the mass turned into stars by the age of the galaxy based on its star formation history. This is an overprediction since during the galaxy's lifetime stars die and leave remnants whose mass is smaller than the initial one \citep{Renziniciotti:93}. For the relatively young age of RCSGA 032727-132609, this mass returned to the interstellar medium (ISM) will be a very small percentage. 

The biggest mass bias is caused by the use of single-component star formation history models. These attribute the emission at all wavelengths from rest-frame UV through near-IR to a single, smoothly declining or continuous episode of star formation, which is not a very realistic picture of the episodic star formation histories that arise in a hierarchical model of galaxy formation. When a young episode of star formation is superimposed on an older burst that peaked sometime in the past, the stellar mass contribution of this older stellar population will likely not be captured by single-component modeling of the overall SED. In order to set a rough upper limit on the range of stellar masses allowed by the SED of RCSGA 032727-132609, we follow \cite{Shapley:05} and \cite{Daddi:04b} in combining a very young model, which dominates the emission at restframe UV wavelengths, with a maximally old underlying burst, which dominates the near-IR emission and whose age is only limited by the age of the Universe at $z=1.701$. Specifically, we scale a young ($t=10$\,Myr) CSF model to the observed $r$-band magnitude, subtract this model from the observed SED, and scale the maximally old ($t=3$\,Gyr, $\tau=100$\,Myr) model to match the residual near-IR magnitudes. The stellar mass is the sum of the two components, largely dominated by the older stellar population, and is found to be about twice as large as our best-fit single-component mass (after correcting for the mass returned to the ISM by supernovae). Similarly, the age inferred from SED modeling represents the duration of the current episode of star formation and could be larger for an underlying older stellar population.

\subsection{Constraints on metallicity}
\label{sec:met}
\begin{figure}[h]
\includegraphics[width=9cm]{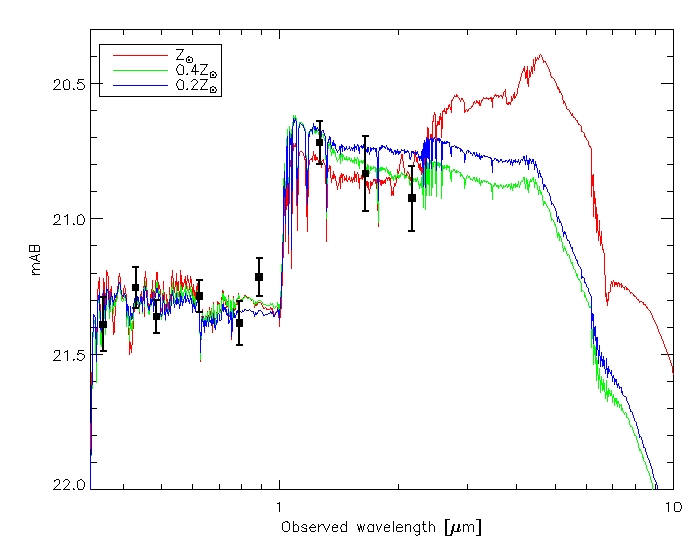}
\vspace{0.05in}
\caption{Best-fit SEDs from Z$_\odot$, 0.4\,Z$_\odot$ and 0.2\,Z$_\odot$ metallicity models for the counter-image. Photometry in $u,B,g,r,Iz,J,H$ and $K_s$-bands is overplotted as black datapoints with $1\sigma$ errorbars.\label{fig:met}}
\end{figure}

We have assumed solar metallicity for the SED models to remove one parameter in the degeneracy between age, dust extinction and metallicity that arises from the spectral slope. It is of interest to explore the effect of sub-solar metallicity models on the best-fit stellar population parameters. The presence of the CIII] nebular emission line could point to a metallicity similar to or less than 0.4\,Z$_\odot$ found for cB58 \citep{Pettini:00}, which does not show this line in its spectrum (see \S~\ref{sec:spec}). We fit the same range of models (Chabrier IMF, Calzetti dust extinction, $\tau = 0.01, 0.05, 0.1, 0.2, 0.5, 1.0, 2.0$\,Gyr or CSF) with 0.4 and 0.2\,Z$_\odot$ metallicities. We find acceptable fits for each metallicity, the best-fit SEDs are shown for the counter-image in Figure~\ref{fig:met}. 

Based on $\chi^2$-statistics, the 0.2\,Z$_\odot$ models produce a worse fit, but the 0.4\,Z$_\odot$ best-fit model reports a better $\chi^2$ of 0.66 and 0.63 for the arc and counter-image respectively, compared to 0.87 and 0.89 for the default solar metallicity. The stellar population parameters from the 0.4\,Z$_\odot$ models are listed in Table~\ref{tab:sed}. They are overall consistent with the confidence intervals deduced earlier from the  solar metallicity models; the smaller ranges allowed for the age and stellar mass confirm that the 0.4\,Z$_\odot$ models produce overall better fits to the observed SED. \cite{Erb:06} have investigated the mass-metallicity relation for galaxies at $z\sim2$ and find a metallicity of $\sim 0.5$\,Z$_\odot$ for a stellar mass of $10^{10}$\,M$_\odot$, consistent with our results. From Figure~\ref{fig:met}, it is obvious that the SEDs start to diverge considerably in the mid-IR region, such that inclusion of IRAC data to the SED could distinguish between these models. This will be explored in a future paper. External metallicity information from spectral emission lines would also be very helpful. 

\begin{deluxetable*}{lccccccc} 
\tablecolumns{8} 
\tablecaption{Stellar population parameters \label{tab:sed}} 
\tablehead{ \multicolumn{1}{c}{} &
	    \multicolumn{2}{c}{ARC}   & 
            \multicolumn{2}{c}{COUNTER-IMAGE}  \\ 
            \multicolumn{1}{c}{} &
	    \multicolumn{1}{c}{Z$_\odot$}   & 
	    \multicolumn{1}{c}{0.4\,Z$_\odot$}   & 
            \multicolumn{1}{c}{Z$_\odot$}   & 
	    \multicolumn{1}{c}{0.4\,Z$_\odot$}} 
\startdata 
Age [Myr]   & 91$\pm$65  & 91$\pm$27  & 143$\pm$90 & 72$\pm$35   \\ 
$E(B-V)$      & $\le0.035$ & $\le0.110$ & $\le0.025$ & $\le0.045$  \\ 
$\tau$ [Myr] & 10-50      & 10-50      & 10-50      & 10-50 \\
$\log(\mathrm{M}/\mathrm{M}_\odot)$ & 10.07$\pm$0.08 & 10.05$\pm$0.07 & 10.01$\pm$0.10 & 9.95$\pm$0.07 \\
$SFR$ [M$_\odot$\,yr$^{-1}$] & $\le35$ & $\le77$ & $\le38$ & $\le41$ \\
\enddata
\end{deluxetable*}

\section{Context}
\label{sec:compall}
\subsection{Luminosity}
\label{sec:lum}
An often quoted measure of the brightness of a galaxy is its comparison to L$_*$, the luminosity of the characteristic break in the Schechter luminosity function \citep{Schechter:76}. There have been a number of recent studies of the luminosity function at $z\sim1.7$. \cite{Arnouts:05} provide an estimate of the UV luminosity function at 1500\AA\ for a sample of galaxies at $1.75 < z < 2.25$ with a characteristic absolute magnitude M$_* = -20.33\pm0.50$. This is consistent with a recent study by \cite{Oesch:10} which reports M$_* = -20.34\pm0.29$ for galaxies at $1.5 < z < 2.0$. We compare this to the absolute magnitude at rest-frame 1500\AA\ of the counter-image of RCSGA 032727-132609, derived from the best fit SED via M$_{1500} = m_{1500} - 5 \log(D_L/10\mathrm{pc}) + 2.5\log(1+z)$. We use the counter-image as the best representative of the source because of its more robust magnification estimate. Taking into account the lensing magnification of $2.04\pm0.16$, we find M$_{1500}=-22.37\pm0.09$. Relative to the characteristic M$_* = -20.34\pm0.29$ postulated by \cite{Oesch:10}, we find RCSGA 032727-132609 at a luminosity of $6.5\pm1.8$\,L$_*$. 

\subsection{Stellar mass}
\label{sec:mass}
Similar to the comparison to L$_*$ made above, we can compare the stellar mass of RCSGA 032727-132609 to M$_{\mathrm{star}}^*$, the mass of the characteristic break in the Schechter mass function \citep{Schechter:76}. \cite{Marchesini:09} present a thorough analysis of the evolution of the stellar mass function from $z \sim 4$ to the present. They devote much attention to the systematic uncertainties induced by the assumptions made in the SED modeling of the stellar masses. For the best-fit Schechter function derived with SED assumptions consistent with ours (CB07, Kroupa IMF-consistent with Chabrier, Calzetti extinction, Z$_\odot$) for the redshift range $1.3 < z < 2.0$, they find $\log(\mathrm{M}_{\mathrm{star}}^*/\mathrm{M}_\odot)=10.80\pm0.05$. When comparing this to the stellar mass of $\log(\mathrm{M}/\mathrm{M}_\odot)=10.01\pm0.10$ obtained from the counter-image, we find RCSGA 032727-132609 at a stellar mass of $0.16\pm0.04$\,M$_{\mathrm{star}}^*$.

\subsection{Comparison to representative galaxy samples}
\label{sec:comp}
In addition to the statistical measures discussed above, we want to place RCSGA 032727-132609 in reference to the overall galaxy population at this redshift that exists in current studies, to determine to what extent the stellar population parameters found in this highly magnified galaxy can be generalized to the known population as a whole. The most representative sample for comparison are the UV selected star-forming galaxies at $1.4 \le z \le 2.5$ collected by Steidel and collaborators \citep{Adelberger:04, Steidel:04, Shapley:05}. They have adapted the color criteria from the succesful LBG continuum break technique at $z\sim3$ (Steidel 1996; 2003) to lower redshifts, defining `BM' and `BX' color criteria for $1.4 \le z \le 2.0$ and $2.0 \le z \le 2.5$ respectively \citep{Adelberger:04}. \cite{Reddy:06a} present SED modeling for a large sample of 212 spectroscopically confirmed `BM/BX' galaxies. Because of the overall faint magnitudes, the star formation histories remain basically unconstrained by SED modeling and they focus on stellar masses as a more robust diagnostic.  Specifically for the subsample of 51 `BM' galaxies at $z = 1.72 \pm 0.34$, they find an average Salpeter stellar mass of $\left<\log(M/M_\odot)\right>=9.86\pm0.57$. Their sample spans a large dynamic range in age (few Myr to 2.5\,Gyr), reddening ($E(B-V) = 0.0-0.5$) and star formation rate ({\it SFR}$ = 0-914$\,M$_\odot$\,yr$^{-1}$).

When fitting the SED of the counter-image of RCSGA 032727-132609 with solar metallicity Salpeter IMF models, we find a stellar mass of $\log(\mathrm{M}/\mathrm{M}_\odot)=10.26\pm0.10$, a factor of 1.8 higher than what was stated above based on the Chabrier IMF, in agreement with the known overprediction of the number of low mass stars by the Salpeter IMF. This is consistent with the average stellar mass found by \cite{Reddy:06a}. The best-fit age, reddening and star formation rate of RCSGA 032727-132609 also fall into the large ranges found for this comparison sample.

\section{Summary}
This paper reports on the discovery of an exceptionally bright and extended star-forming galaxy at $z=1.701$, strongly-lensed by a foreground cluster discovered in the RCS2 survey. The giant arc is $\sim 3.5$ times brighter than cB58 and extends over 38\,\arcsec on the sky. We measure a velocity dispersion of 988$\pm$122\,km~s$^{-1}$ for the cluster and estimate a virial mass of $M_{200} \sim 1.1 \times 10^{15}$\,M$_\odot$\,$h_{70}^{-1}$, in accord with the large Einstein-radius of $\sim 17\farcs8$. A lens model is constructed for the cluster using the publicly-available software {\tt LENSTOOL} \citep{Jullo:07} and results in a magnification factor of 2.04$\pm$0.16 for the counter-image and an estimate of 17.2$\pm$1.4 for the average magnification of the giant arc, based on the relative sizes of the arc and the counter-image on the sky. Higher resolution imaging is required to create a robust lens model for the giant arc which correctly explains the apparent substructure. 

Careful measurements of consistent magnitudes for 9 bands of photometry ranging from $u$ to $K_s$ produce a well-constrained spectral energy distribution. SED fitting is based on CB07 models with a Chabrier IMF and Calzetti dust extinction. The systematic uncertainties which arise from the choice of stellar evolution model, IMF and extinction law can be large and often exceed the statistical uncertainties from photometric errors. Our best-fit model places RCSGA 032727-132609 at a metallicity of 0.4\,Z$_\odot$ with a moderately young age, $t=80\pm40$\,Myr and a relatively small amount of dust extinction, $E(B-V) \le 0.11$. Taking into account the lensing magnification, we find a stellar mass of M$_* \sim 10^{10}$\,M$_\odot$ and a current star formation rate $SFR \le 77$\,M$_\odot$\,yr$^{-1}$. The agreement between the stellar population parameters independently inferred from the giant arc and counter-image, both images of the same background galaxy, is a good consistency check of the photometry, SED fitting procedure and lens modeling. Single-component SED models report ages and stellar mass measurements for the current episode of star formation. When allowing for an episodic star formation history, twice this mass could be present in an older underlying burst without affecting the SED detectably. 

A comparison to the characteristic L$_*$ and M$_{\mathrm{star}}^*$ of the Schechter luminosity and mass functions at $z\sim1.7$ places RCSGA 032727-132609 at a luminosity of $6.5\pm1.8$\,L$_*$ and a stellar mass of $0.16\pm0.04$\,M$_{\mathrm{star}}^*$, suggesting a low-mass starbursting galaxy. The Salpeter stellar mass estimate of RCSGA 032727-132609 is consistent with the average stellar mass for a sample of `BM' galaxies at $z = 1.72 \pm 0.34$ \citep{Reddy:06a} and its age, reddening and current star formation rate generally fall into the broad range of stellar population parameters found for this comparison sample. The large magnification and spatial extent of RCSGA 032727-132609 provide a unique opportunity to spatially resolve one representative example of the variety of stellar populations found at these redshifts and study it in great detail. This paper has presented only our first level of understanding of this unique galaxy.

\begin{acknowledgments}
We thank the anonymous referee for a careful reading of the manuscript and helpful comments. LFB research is partially funded by Centro de Astrofisica FONDAP and by proyecto FONDECYT 1085286. RPM acknowledges support from a Conicyt Doctoral fellowship, Programa MECESUP 2 c\'odigo PUC0609, and Fondo Gemini 2008 proyecto No. 32080004.
\end{acknowledgments}


\begin{thebibliography}{}

\bibitem[Adelberger et al.(2004)]{Adelberger:04}
{Adelberger}, K.~L., {Steidel}, C.~C., {Shapley}, A.~E., {Hunt}, M.~P., {Erb}, D.~K., {Reddy}, N.~A., {Pettini}, M. 2004, \apj, 607, 226

\bibitem[Allam et al.(2007)]{Allam:07}
{Allam}, S.~S., {Tucker}, D.~L., {Lin}, H., {Diehl}, H.~T., {Annis}, J., {Buckley-Geer}, E.~J., {Frieman}, J.~A. 2007, \apjl, 662, L51


\bibitem[Arnouts et al.(2005)]{Arnouts:05}
{Arnouts}, S. et al. 2005, \apjl, 619, L43

\bibitem[Blain et al.(1999)]{Blain:99}
{Blain}, A.~W., {Smail}, I., {Ivison}, R.~J., {Kneib}, {J.-P.} 1999, \mnras, 302, 632

\bibitem[Bolzonella et al.(2000)]{Bolzonella:00}
{Bolzonella}, M., {Miralles}, {J.-M.}, {Pell{\'o}}, R. 2000, \aap, 363, 476
 
 

\bibitem[Bouwens et al. (2009)]{Bouwens:09}
{Bouwens}, R.~J. et al. 2009, \apj, 690, 1764

\bibitem[Calzetti et al.(2000)]{Calzetti:00}
{Calzetti}, D., {Armus}, L., {Bohlin}, R.~., {Kinney}, A.~L., {Koornneef}, J., {Storchi-Bergmann}, T. 2000, \apj, 533, 682
 
\bibitem[Chabrier(2003)]{Chabrier:03}
{Chabrier}, G. 2003, \pasp, 115, 763
 
\bibitem[Chapman et al.(2003)]{Chapman:03}
{Chapman}, S.~C., {Blain}, A.~W., {Ivison}, R.~J., {Smail}, I.~R. 2003, \nat, 422, 695

\bibitem[Chonis \& Gaskell(2008)]{Chonisgaskell:08}
{Chonis}, T.~S., {Gaskell}, C.~M. 2008, \aj, 135, 264

 
\bibitem[Conroy \& Gunn (2010)]{Conroy:09}
{Conroy}, C., {Gunn}, J.~E. 2010, \apj, 712, 833

\bibitem[Coppin et al.(2007)]{Coppin:07}
{Coppin}, K.~E.~K. et al. 2007, \apj, 665, 936

\bibitem[Daddi et al.(2004a)]{Daddi:04a}
{Daddi}, E. et al. 2004, \apj, 617, 746

\bibitem[Daddi et al.(2004b)]{Daddi:04b}
{Daddi}, E. et al. 2004, \apjl, 600, L127
 
  
\bibitem[Ellingson et al.(1996)]{Ellingson:96}
{Ellingson}, E., {Yee}, H.~K.~C., {Bechtold}, J., {Elston}, R. 1996, \apjl, 466, L71

\bibitem[Erb et al.(2006)]{Erb:06}
{Erb}, D.~K., {Shapley}, A.~E., {Pettini}, M., {Steidel}, C.~C., {Reddy}, N.~A., {Adelberger}, K.~L. 2006, \apj, 644, 813

\bibitem[Finkelstein et al.(2009)]{Finkelstein:09}
{Finkelstein}, S.~L., {Cohen}, S.~H., {Malhotra}, S., {Rhoads}, J.~E. 2009, \apj, 700, 276
 
\bibitem[Gladders \& Yee(2000)]{Gladdersyee:00}
{Gladders}, M.~D., {Yee}, H.~K.~C. 2000, \aj, 120, 2148
  
\bibitem[Halliday et al.(2008)]{Halliday:08}
{Halliday}, C. et al. 2008, \aap, 479, 417

\bibitem[Heckman et al.(1998)]{Heckman:98}
{Heckman}, T.~M., {Robert}, C., {Leitherer}, C., {Garnett}, D.~R., {van der Rydt}, F. 1998, \apj, 503, 646

\bibitem[Jordi et al.(2006)]{Jordi:06}
{Jordi}, K., {Grebel}, E.~K., {Ammon}, K. 2006, \aap, 460, 339

\bibitem[Jullo et al.(2007)]{Jullo:07}
{Jullo}, E., {Kneib}, {J.-P.}, {Limousin}, M., {El{\'{\i}}asd{\'o}ttir}, {\'A}., {Marshall}, P.~J., {Verdugo}, T. 2007, ArXiv e-prints, arXiv0706.0048

\bibitem[Koester et al.(2010)]{Koester:10}
{Koester}, B.~P. et al. 2010, ArXiv e-prints, arXiv1003.0030

\bibitem[Kriek et al.(2008)]{Kriek:08}
{Kriek}, M. et al. 2008, \apj, 677, 219

\bibitem[Limousin et al.(2005)]{Limousin:05}
{Limousin}, M., {Kneib}, {J.-P.}, {Natarajan}, P. 2005, \mnras, 356, 309

\bibitem[Limousin et al.(2007)]{Limousin:07}
{Limousin}, M. et al. 2007, \apj, 668, 643


\bibitem[Maraston(2005)]{Maraston:05}
{Maraston}, C. 2005, \mnras, 362, 799

\bibitem[Marchesini et al.(2009)]{Marchesini:09}
{Marchesini}, D., {van Dokkum}, P.~G., {F{\"o}rster Schreiber}, N.~M., {Franx}, M., {Labb{\'e}}, I., {Wuyts}, S. 2009, \apj, 701, 1765

\bibitem[Muzzin et al.(2009)]{Muzzin:09}
{Muzzin}, A., {Marchesini}, D., {van Dokkum}, P.~G., {Labb{\'e}}, I., {Kriek}, M., {Franx}, M. 2009, \apj, 701, 1839

\bibitem[Oesch et al.(2010)]{Oesch:10}
{Oesch}, P.~A. et al. 2010, arXiv1005.1661

\bibitem[Papovich et al.(2001)]{Papovich:01}
{Papovich}, C., {Dickinson}, M., {Ferguson}, H.~C. 2001, \apj, 559, 620
   
\bibitem[Peng et al.(2002)]{Peng:02}
{Peng}, C.~Y., {Ho}, L.~C., {Impey}, C.~D., {Rix}, {H.-W.} 2002, \aj, 124, 266
 
\bibitem[Pettini et al.(2000)]{Pettini:00}
{Pettini}, M., {Steidel}, C.~C., {Adelberger}, K.~L., {Dickinson}, M., {Giavalisco}, M. 2000, \apj, 528, 96

\bibitem[Pettini et al.(2002)]{Pettini:02}
{Pettini}, M., {Rix}, S.~A., {Steidel}, C.~C., {Adelberger}, K.~L., {Hunt}, M.~P., {Shapley}, A.~E. 2002, \apj, 569, 742
  

\bibitem[Reddy et al.(2006a)]{Reddy:06a}
{Reddy}, N.~A., {Steidel}, C.~C., {Erb}, D.~K., {Shapley}, A.~E., {Pettini}, M. 2006, \apj, 653,1004
  
\bibitem[Reddy et al.(2006b)]{Reddy:06b}
{Reddy}, N.~A. et al. 2006, \apj, 644, 792


\bibitem[Reddy \& Steidel(2009)]{Reddy:09}
{Reddy}, N.~A., {Steidel}, C.~C. 2009, \apj, 692, 778

\bibitem[Reddy et al.(2010)]{Reddy:10}
{Reddy}, N.~A., {Erb}, D.~K., {Pettini}, M., {Steidel}, C.~C., {Shapley}, A.~E. 2010, \apj, 712, 1070
  
\bibitem[Renzini \& Ciotti(1993)]{Renziniciotti:93}
{Renzini}, A., {Ciotti}, L. 1993, \apjl, 416, L49
 
\bibitem[Richard et al.(2008)]{Richard:08}
{Richard}, J., {Stark}, D.~P., {Ellis}, R.~S., {George}, M.~R., {Egami}, E., {Kneib}, J.-P., {Smith}, G.~P. 2008, \apj, 685, 705

\bibitem[Salpeter(1955)]{Salpeter:55}
{Salpeter}, E.~E. 1955, \apj, 121, 161


\bibitem[Schechter (1976)]{Schechter:76}
{Schechter}, P. 1976, \apj, 203, 297

\bibitem[Schlegel (1998)]{Schlegel:98}
{Schlegel}, D.~J., {Finkbeiner}, D.~P., {Davis}, M. 1998, \apj, 500, 525

\bibitem[Seitz et al.(1998)]{Seitz:98}
 {Seitz}, S., {Saglia}, R.~P., {Bender}, R., {Hopp}, U., {Belloni}, P., {Ziegler}, B. 1998, \mnras, 298, 945

\bibitem[Shapley et al.(2003)]{Shapley:03}
{Shapley}, A.~E., {Steidel}, C.~C., {Pettini}, M., {Adelberger}, K.~L. 2003, \apj, 588, 65
 
 
\bibitem[Shapley et al.(2005)]{Shapley:05}
{Shapley}, A.~E. et al. 2005, \apj, 626, 698
 
\bibitem[Siana et al.(2008)]{Siana:08}
{Siana}, B., {Teplitz}, H.~I., {Chary}, {R.-R.}, {Colbert}, J., {Frayer}, D.~T. 2008, \apj, 689, 59

\bibitem[Siana et al.(2009)]{Siana:09}
{Siana}, B. et al. 2009, \apj, 698, 1273

\bibitem[Smail et al.(2007)]{Smail:07}
{Smail}, I. et al. 2007, \apjl, 654, L33

\bibitem[Steidel et al.(1996)]{Steidel:96}
{Steidel}, C.~C., {Giavalisco}, M., {Pettini}, M., {Dickinson}, M., {Adelberger}, K.~L. 1996, \apj, 462L, 17S

\bibitem[Steidel et al.(2003)]{Steidel:03}
{Steidel}, C.~C., {Adelberger}, K.~L., {Shapley}, A.~E., {Pettini}, M., {Dickinson}, M., {Giavalisco}, M. 2003, \apj, 592, 728

\bibitem[Steidel et al.(2004)]{Steidel:04}
{Steidel}, C.~C., {Shapley}, A.~E., {Pettini}, M., {Adelberger}, K.~L., {Erb}, D.~K., {Reddy}, N.~A., {Hunt}, M.~P. 2004, \apj, 604, 534


\bibitem[Williams \& Lewis(1996)]{Williams:96}
{Williams}, L.~L.~R., {Lewis}, G.~F. 1996, \mnras, 281, L35
 
\bibitem[Wuyts et al.(2009)]{Wuyts:09}
{Wuyts}, S., {Franx}, M., {Cox}, T.~J., {Hernquist}, L., {Hopkins}, P.~F., {Robertson}, B.~E., {van Dokkum}, P.~G. 2009, \apj, 696, 348   

\bibitem[Yee et al.(1996)]{Yee:96}
{Yee}, H.~K.~C., {Ellingson}, E., {Bechtold}, J., {Carlberg}, R.~G., {Cuillandre}, {J.-C.} 1996, \aj, 111, 1783
    
\bibitem[Yee \& Ellingson(2003)]{yee03}
Yee, H.~K.~C., Ellingson, E. 2003, \apj, 585, 215

\end{thebibliography}
\end{document}